\documentclass[12pt,a4paper]{article}
\usepackage{amsfonts,latexsym}
\usepackage{graphicx,color}
\usepackage{dcolumn}
\usepackage{graphicx}
\usepackage{amsmath}
\usepackage{amsfonts}
\usepackage{amssymb}
\usepackage{ulem}

\renewcommand{\thefootnote}{\fnsymbol{footnote}}

\begin{document}

\vspace{12mm}

\begin{center}
{{{\Large {\bf Scalarized extremal black holes in the Einstein-Maxwell-scalar theory with two U(1) fields}}}}\\[10mm]
{Xiao Yan Chew$^1$\footnote{e-mail address: xiao.yan.chew@just.edu.cn} and Yun Soo Myung$^{2}$\footnote{e-mail address: ysmyung@inje.ac.kr}}\\[8mm]

{${}^1$School of Science, Jiangsu University of Science and Technology, Zhenjiang, 212100 China\\[0pt]}
{${}^2$ Center for Quantum Spacetime, Sogang University, Seoul 04107, Republic of Korea\\[0pt]}

\end{center}
\vspace{2mm}

\begin{abstract}
We study  scalarized extremal  black holes  in  the Einstein-Maxwell-scalar (EMS) theory with two different scalar couplings to  two U(1) fields.
This theory is inspired  by the dyonic EMS theory.
Two scalarzied extremal black holes are found with constant scalar hair.
We confirm that these are exactly  obtained from the extremal  scalarization and  entropy function approaches.
This may imply that it is not easy to find extremal black holes with primary scalar hair.

\end{abstract}
\vspace{5mm}

\vspace{1.5cm}

\hspace{11.5cm}
\newpage
\renewcommand{\thefootnote}{\arabic{footnote}}
\setcounter{footnote}{0}


\section{Introduction}
First of all, we would like to  mention that the black holes with primary scalar hair were found from the (beyond) Horndeski theory where the shift symmetry of the scalar ($\phi\to \phi+$ const.) exists~\cite{Babichev:2013cya,Herdeiro:2015waa,Bakopoulos:2023fmv}. In this case, the primary scalar charge is  an independent one and it  is related to the shift-symmetric charge. 

 A nonminimal scalar coupling to the Gauss-Bonnet (GB) curvature  has induced  the instability of Schwarzschild  black holes and thus, led to spontaneous scalarization triggered by tachyonic scalar~\cite{Doneva:2017bvd,Silva:2017uqg,Antoniou:2017acq}. This is  a modern way to obtain  black holes with secondary scalar hair  and it  plays a key role to understand the interaction between gravity and scalar~\cite{Doneva:2022ewd}. Here, one usually finds secondary scalar charges because spontaneous scalarization requires 
 the regularity of horizon scalar and the asymptotic condition, leading to being dependent on the known hairs and action variables.  
Furthermore, a  nonminimal coupling to the Maxwell invariant  has  demonstrated  spontaneous scalarization of Reissner-Nodstr\"{o}m (RN) black holes in the Einstein-Maxwell-scalar (EMS) theory. It led  to scalarized non-extremal  black holes with secondary scalar hair~\cite{Herdeiro:2018wub} because the spontaneous scalarization needs to impose the asymptotic condition.

The comparative  analysis for dilaton  and  scalar couplings  has revealed  that two cases  have provided charged black holes with secondary hair~\cite{Astefanesei:2019pfq}.
 A difference is that  the former did not accommodate any extremal black holes with scalar hair, while the latter was   allowed to possess  scalarized extremal  black holes  when extending to a dyon with electric and magnetic charges. However, we  wish to point out that  their extremal black hole solution  with secondary scalar hair does not seem  to be a promising  one because the  extremal condition  for a metric function ($\delta(r)=0$) is not imposed on finding the scalarized extremal black holes.

Up to now, one believes  that  finding a scalarized extremal black hole with secondary  scalar hair is considered as  a little difficult task analytically and numerically.
It is clear that one known solution is the  Bocharova–Bronnikov–Melnikov–Bekenstein (BBMB) black hole found from the Eintein-conformally coupled scalar (ECS) theory~\cite{Bocharova:1970skc,Bekenstein:1974sf}.
It took a  compact form of the secondary scalar hair  $\phi(r)=m/(r-m)$ with the  black hole mass $m$, even though it blows up at the horizon~\cite{Charmousis:2015aya}.
It is worth noting   that the BBMB black hole is regarded as the unique static and asymptotically
flat solution to the ECS theory, evading no hair theorem~\cite{Xanthopoulos:1992fm}.
Furthermore, this BBMB solution was recovered in a nontrivial way from the numerical series solution when imposing  the asymptotically flat condition~\cite{Myung:2019adj,Zou:2019ays}. Before imposing this condition, the scalar charge was  primary. 
 
On the other hand,  spontaneous scalarization of RN black holes was investigated when approaching  to extremality by considering the Einstein-Maxwell-Gauss-Bonnet-scalar theory with a quadratic scalar coupling to GB term~\cite{Brihaye:2019kvj}. Two branches of scalarized black holes have appeared  when approaching to extremality when using the Bertotti-Robinson  geometry (AdS$_2\times S^2$) as the near-horizon approximation to the extremal black hole.
Also, there exist some examples of entropy function approach~\cite{Sen:2005wa} to find scalarized extremal black holes~\cite{Marrani:2017uli,Astefanesei:2019pfq,Marrani:2022hva,Myung:2026ook}.
At this stage, we wish to point out  its limitation: this approach  is not suitable for  finding  extremal black holes with secondary scalar hair because it uses an attractor mechanism with constant scalar existing   on the degenerate horizon.

In this work, we will introduce the  EMS theory with two different scalar couplings to two  U(1) fields  to find  scalarized extremal black holes.
This action  is closely related to  the bosonic sector of $N=4$ supergravity which has admitted the dilatonic black hole located at $r=r_+=0$ with secondary dilaton hair. Unfortunately, it  did  not include  an extremal black hole with any dilaton hair~\cite{Kol:1996hf,Krasnitz:1997gn}.
We note that this model was also used to explore spontaneous scalarization of non-extremal (RN)  black hole by choosing exponential  coupling function $f(\phi)=e^{\alpha \phi^2}$~\cite{Myung:2020dqt}.
Actually,  this model  is very similar to the dyonic EMS theory which was used to find  scalarized extremal black holes~\cite{Astefanesei:2019pfq,Chen:2026olq}.
We find two scalarzied extremal black holes with constant scalar hair. It is worthy to note that these solutions  are exactly  recovered from  both extremal  scalarization  and  entropy function approaches.

\section{EMS theory with U(1) fields  } \label{sec1}
First of all, we mention briefly the bosonic sector for $N=4$ supergravity~\cite{Kol:1996hf,Krasnitz:1997gn,Lee:1997xg}
\begin{equation}
S_{\rm N4}=\frac{1}{16 \pi}\int d^4 x\sqrt{-g}\Big[ R-2\partial_\mu \phi \partial^\mu \phi-e^{-2 \phi} F^2-e^{2 \phi} H^2\Big],\label{act1}
\end{equation}
where $\phi$ is the  dilaton,   $F=dA$ and $H=dB$ correspond to  two U(1) field strengths.
An analytic black hole solution was  found to be 
\begin{equation}
ds^2_{\rm N4}=-\frac{dt^2}{H_1H_2}+H_1H_2\Big(dr^2+r^2d\Omega^2_2\Big),\quad e^{2\bar{\phi}}=\frac{H_2}{H_1},~ \bar{F}=\frac{1}{\sqrt{2}}dH_1^{-1}\wedge dt,~\bar{H}=\frac{1}{\sqrt{2}}dH_2^{-1}\wedge dt
\end{equation}
with two harmonic  functions $H_1=1+\frac{\sqrt{2}Q}{r}$ and $H_2=1+\frac{\sqrt{2}P}{r}$.
The event horizon is unfortunately  located at $r=r_{+}=0$. The dilaton $\bar{\phi}$ is given by 
the secondary hair $\bar{\phi}=\frac{1}{2}\ln[H_2(r)/H_1(r)]$ and its horizon dilaton takes the form 
$\lim_{r\to 0}\bar{\phi}=\frac{1}{2}\ln[\frac{P}{Q}]$.
In case of $P=Q$ (becoming extremal black hole), however, one finds that $\bar{\phi}=0$.  This implies that the extremal black hole could not have any dilaton hair. This is not a desired case.

Hence, we introduce  a new action of  the EMS theory with two U(1) fields by replacing the dilaton coupling $ e^{2 \phi}$ with  a scalar coupling function  $f(\phi)$ in Eq.(\ref{act1})
\begin{equation}
S_{\rm EMS}=\frac{1}{16 \pi}\int d^4 x\sqrt{-g}\Big[ R-2\partial_\mu \phi \partial^\mu \phi-\frac{ F^2}{f(\phi)}-f(\phi) H^2\Big].\label{cct2}
\end{equation}
Then, the Einstein  equation is derived as
\begin{eqnarray}
 G_{\mu\nu}=2\partial _\mu \phi\partial _\nu \phi -(\partial \phi)^2g_{\mu\nu}+2T^{U(1)}_{\mu\nu}, \label{equa1}
\end{eqnarray}
where
\begin{equation}
T^{U(1)}_{\mu\nu}=\frac{1}{f(\phi)}\Big(F_{\mu\rho}F_{\nu}~^\rho-\frac{F^2}{4}g_{\mu\nu}\Big)+f(\phi)\Big(H_{\mu\rho}H_{\nu}~^\rho-\frac{H^2}{4}g_{\mu\nu}\Big).
\end{equation}
Two  Maxwell equations are given by
\begin{eqnarray} \label{M-eq}
&&\partial_\mu \Big(\sqrt{-g}\frac{F^{\mu\nu}}{f(\phi)}\Big)=0,\quad \partial_\mu \Big(\sqrt{-g}f(\phi)H^{\mu\nu}\Big)=0.
\end{eqnarray}
Finally, the scalar equation takes the form
\begin{equation}
\square \phi +\frac{f'(\phi)}{4f(\phi)}\Big(-\frac{F^2}{f(\phi)}+f(\phi) H^2\Big)=0 \label{s-equa}.
\end{equation}

\section{Scalarized extremal black holes}

To obtain scalarized  black holes, we  introduce  the metric and fields as~\cite{Herdeiro:2018wub}
\begin{eqnarray}\label{nansatz}
ds^2_{\rm SCBH}&=&\bar{g}_{\mu\nu}dx^\mu dx^\nu=-N(r)e^{-2\delta(r)}dt^2+\frac{dr^2}{N(r)}+r^2(d\theta^2+\sin^2\theta d\varphi^2), \nonumber \\
N(r)&=&1-\frac{2m(r)}{r},\quad\bar{\phi}= \phi(r),\quad \bar{A}_t=v_Q(r),\quad \bar{B}_t=v_P(r).
\end{eqnarray}
Substituting Eq.(\ref{nansatz}) into Eqs.(\ref{equa1})-(\ref{s-equa}) with $v'_Q=\frac{e^{-\delta}Qf(\phi)}{r^2}$ and $v'_P=\frac{e^{-\delta}P}{r^2f(\phi)}$, one has three  ordinary differential equations
\begin{eqnarray}
&&m'(r)=\frac{1}{2} r^2 N \phi'^2(r)+\frac{1}{2r^2}\left(f(\phi)Q^2+\frac{P^2}{f(\phi)} \right),\label{eqq-1} \\
&&\delta'(r)=-r \phi'^2(r), \label{eqq-2} \\
&& (e^{-\delta} r^2 N\phi'(r))'+\frac{ e^{-\delta}}{2r^2}
\frac{ f'(\phi)}{ f(\phi)} \left(f(\phi)Q^2-\frac{P^2}{f(\phi)} \right)=0, \label{eqq-3}
\end{eqnarray}
where the prime ($'$) denotes derivative with respect to  its argument. It is desirable to note that the second term of Eq.(\ref{eqq-3}) gives a hint to obtain a constant scalar solution  when imposing either $f'(\phi)=0$ or $f(\phi)=\frac{P}{Q}$.

Here, we  could obtain three non-extremal and extremal  black hole solutions depending the scalar hair $\phi$.
Firstly, one finds the dyonic black hole and its extremal black hole for $\phi=0$ and $f(0)=1$:
\begin{equation}
N(r)=1-\frac{2M}{r}+\frac{Q^2+P^2}{r^2},\quad \delta(r)=0, \quad N_e(r)=\Big(1-\frac{M}{r}\Big)^2 \quad {\rm for }\quad M=\sqrt{Q^2+P^2}. \label{Cbh-1}
\end{equation}
Secondly, we obtain one scalarized non-extremal  black hole and its extremal black hole for $\phi=\phi_c$ and $f(\phi_c)=\frac{P}{Q}$:
\begin{equation}
N_{a}(r)=1-\frac{2M}{r}+\frac{2QP}{r^2},\quad \delta(r)=0, \quad N_{ae}(r)=\Big(1-\frac{M}{r}\Big)^2 \quad {\rm for }\quad M=\sqrt{2QP}=r_e. \label{Cbh-2}
\end{equation}
Finally, the other scalarized  non-extrmal  black hole and its extremal black hole are found for $\phi=\tilde{\phi}_c$ and $f'(\tilde{\phi}_c)=0$:
\begin{eqnarray}
N_{b}(r)&=&1-\frac{2M}{r}+\frac{Q^2 f(\tilde{\phi}_c)+\frac{P^2}{f(\tilde{\phi}_c)}}{r^2},\quad \delta(r)=0,\nonumber \\
N_{be}(r)&=&\Big(1-\frac{M}{r}\Big)^2 \quad {\rm for }\quad M=\sqrt{Q^2 f(\tilde{\phi}_c)+\frac{P^2}{f(\tilde{\phi}_c)}}\equiv r_e. \label{Cbh-3}
\end{eqnarray}
We note that for $Q=P$, the second solution becomes  the first one  without scalar hair.
Also, the extremal condition of $\delta(r)=0$ plays an important role in obtaining  scalarized (extremal) black holes with constant scalar hair.

\section{Extremal  scalarization approach}

Before we proceed, we would like to mention that spontaneous scalarization of non-extremal black holes was performed with exponential coupling $e^{\alpha \phi^2}$~\cite{Myung:2020dqt}, where the tachyonic scalar was used to trigger scalarization. In this case, one has found infinite branches of scalarized non-extremal black holes.

In this section, we use the extremal scalarization approach to find scalarized extremal black holes.
For this purpose, we need to introduce a constraint equation as~\cite{Astefanesei:2019pfq}
\begin{equation}
\frac{N''}{2}-N\delta''+N'\Big(\frac{1}{r}-\frac{3\delta'}{2}\Big)+N\delta'\Big(\delta'-\frac{1}{r}\Big)+N\phi'^2=\frac{1}{r^4}\Big(f(\phi)Q^2+\frac{P^2}{f(\phi)}\Big),\label{cons-eq}
\end{equation}
which was  obtained  from combining  Eqs.(\ref{eqq-1})-(\ref{eqq-3}) with  their first derivatives.
In this approach, it is convenient to express Eq.(\ref{eqq-1})  in terms of the  metric function  $N$  by exploiting a relation of  $m'(r)=\frac{1-N-rN'}{2}$.

Now, we consider the near-horizon forms which are suitable for scalarzied extremal black holes
\begin{eqnarray}
&&N(r)=N_2(r-r_+)^2+N_3(r-r_+)^3\cdots,\quad
\delta(r)=\delta_0+\delta_1(r-r_+)+\cdots,\label{aps-1}\\
&&v_{Q}(r)=v_{Q1}(r-r_+)+\cdots,\quad \quad
v_{P}(r)=v_{P1}(r-r_+)+\cdots,\label{aps-2}\\
&&\phi(r)=\phi_0+\phi_1(r-r_+)+\cdots,\label{aps-3}
\end{eqnarray}
where
\begin{eqnarray}
&& \quad v_{Q1}=\frac{e^{-\delta_0}Qf(\phi_0)}{r_+^2},\quad  v_{P1}=\frac{e^{-\delta_0}P}{r_+^2f(\phi_0)}.  \label{eqn-co}
\end{eqnarray}
We note that $\delta_0$ is considered as a free parameter and $\phi_0$ denotes the horizon  scalar.
Two  Eqs.(\ref{eqq-1}) and (\ref{eqq-2}) determine, respectively,
\begin{equation}
r_+^2=f(\phi_0)Q^2+\frac{P^2}{f(\phi_0)},\quad \delta_1=-r_+\phi_1^2. \label{loweq-1}
\end{equation}
On the other hand,  Eq.(\ref{eqq-3}) implies
\begin{equation}
 f'(\phi_0)=0,\quad f(\phi_0)=\frac{P}{Q}, \label{loweq-2}
\end{equation}
while its first  derivative  indicates $\phi_1$-equation as 
\begin{eqnarray}
&&\frac{f'(\phi_0)}{2r_+}\Big(Q^2-\frac{P^2}{f^2(\phi_0)}\Big)\phi_1^2+ \nonumber \\
&&\Big[2r_+^2N_2+\frac{f''(\phi_0)}{2r_+^2}\Big(Q^2-\frac{P^2}{f^2(\phi_0)}\Big)+\frac{f'^2(\phi_0)P^2}{r_+^2 f^3(\phi_0)}\Big]\phi_1-\frac{f'^2(\phi_0)}{r_+^3}\Big(Q^2-\frac{P^2}{f^2(\phi_0)}\Big)=0.\label{phi1-eq}
\end{eqnarray}
Finally,  Eq.(\ref{cons-eq}) leads to
\begin{equation}
 N_2=\frac{1}{r_+^4}\Big(f(\phi_0)Q^2+\frac{P^2}{f(\phi_0)}\Big). \label{loweq-3}
\end{equation}
For $f'(\phi_0)\not=0$ and $N_2\not=-\frac{f'^2(\phi_0)P^2}{2r_+^4f^3(\phi_0)}$, Eqs.(\ref{loweq-1})-(\ref{loweq-3}) determine
\begin{equation}
f(\phi_0)=\frac{P}{Q}, \quad r_+=\sqrt{2QP},\quad  \phi_1=\delta_1=0, \quad N_2=\frac{1}{r_+^2}. \label{Nff1}
\end{equation}
Also, for $f'(\phi_0)=0$ $(f(\phi_0)\not=P/Q)$, Eqs.(\ref{loweq-1})-(\ref{loweq-3}) imply unless $N_2=-\frac{f''(\phi_0)}{4r_+^4}\Big(Q^2-\frac{P^2}{f^2(\phi_0)}\Big)$
\begin{equation}
r_+=\sqrt{f(\phi_0)Q^2+\frac{P^2}{f(\phi_0)}},\quad \phi_1=\delta_1=0,\quad  N_2=\frac{1}{r_+^2}. \label{Nff2}
\end{equation}

At this stage, one has to match Eqs.(\ref{aps-1})-(\ref{aps-3}) with  the asymptotic forms in the far-region
\begin{eqnarray}\label{ncoe-a}
&&m(r)=M-\frac{f(\phi_\infty)Q^2+P^2/f(\phi_\infty)+Q_s^2}{2r}+\cdots,\quad
\delta(r)=\frac{Q_s^2}{2r^2}+\cdots,\nonumber\\
&&v_{Q}(r)=\Phi_Q-\frac{Q}{r}+\cdots,\qquad \qquad \qquad
v_{P}(r)=\Phi_P-\frac{P}{r}+\cdots,\nonumber\\
&&\phi(r)=\phi_\infty+\frac{Q_s}{r}+\cdots,
\end{eqnarray}
where  $Q_s$, $\Phi_Q$, $\Phi_P$, and  $\phi_\infty$ denote the scalar charge, the electrostatic potentials at infinity, and the scalar field at infinity, in addition to the ADM mass $M$  and two electric charges ($Q,P$).
It is reseonable to   choose $\phi_\infty=\phi_0$.
We stress   that the  key ingredient  to obtain  extremal black hole with scalar hair is to  impose  an extremal  condition of  $\delta(r)= 0$.
This suggests strongly  that $\delta_0=\delta_1=0$ and $Q_s=0$.

For any $f'(\phi_0)\not=0$, considering Eqs.(\ref{Nff1}) and (\ref{ncoe-a}),
one recovers scalarized extremal black hole Eq.(\ref{Cbh-2}) exactly when choosing $\phi_\infty=\phi_0=\phi_c$.

For $f'(\phi_0)=0$, taking into account  Eqs.(\ref{Nff2}) and (\ref{ncoe-a}), we find scalarized extremal black hole Eq.(\ref{Cbh-3})  for  $\phi_\infty=\phi_0=\tilde{\phi}_c$.

Finally, we  wish to comment  that  the  scalarized extremal black hole solution found in~\cite{Astefanesei:2019pfq} does not seem  to be a promising  one because the extremal  condition for a metric function ($\delta(r)=0$) might not be imposed on finding the scalarized extremal black holes.
If this condition is further imposed, their solution may lead to our solutions.

\section{Entropy function approach}

The previous  construction of scalarized  extremal black holes has showed  extremal black hole with constant scalar hair.
In this section, we wish to adopt the entropy function approach to finding  scalarized  extremal black holes by considering Bertotti-Robinson  geometry  AdS$_2\times S^2$ as the near-horizon approximation of extremal black hole.
For this purpose, we introduce the line element with two unknown parameters $v_0$ and $v_1$ as 
\begin{equation}
\label{AdS2S2}
  ds^2_{\rm BR}=v_0\left(-r^2dt^2+\frac{dr^2}{r^2}\right)+v_1(d\theta^2+\sin^2\theta d\varphi^2) 
\end{equation}
as well as  the matter fields ansatz on the horizon as 
\begin{equation}
\label{matter-attractors}
\phi=\phi_0,\quad A=e r dt,\quad  B=p r dt.
\end{equation}
Here, we note that there is no room  to accommodate  $\delta(r)\not=0$ when considering  the  near-horizon approximation of the extremal black hole. Hence, we may  choose $\delta(r)=0$ as an extremal condition.
Five  parameters
$\{v_0,v_1,\phi_0, e,p\}$
satisfy a set of algebraic relations which result from
Eqs.(\ref{equa1}), (\ref{M-eq}), and (\ref{s-equa}). Here, instead of attempting to solve these, we wish to determine these parameters
by making use of  the entropy function approach~\cite{Sen:2005wa,Astefanesei:2007bf,Sen:2007qy}.
This approach allows us also to compute the black hole entropy which is considered as  the only physical  quantity to describe scalarized extremal black holes.

For this purpose, we introduce the Lagangian density
\begin{eqnarray}
  {\cal L} &=& \frac{1}{16 \pi}\int d\theta d\varphi\sqrt{-g}
	\Big[ R-2\partial_\mu \phi \partial^\mu \phi-\frac{ F^2}{f(\phi)}-f(\phi) H^2 \Big] \nonumber  \\
&=&\frac{1}{2}
\left[
 v_0-v_1+\frac{ v_1}{v_0}\left(\frac{e^2}{f(\phi)}+f(\phi)p^2\right)
\right]. \label{act-E}
\end{eqnarray}
Now, we are in a position to  define the entropy function $\mathcal{E}$
by taking the Legendre transform
of the above density with respect to two electric charges $Q$ and $P$ as
\begin{eqnarray}
\mathcal{E}=
  2\pi \Big(e Q +p P-{\cal L}\Big).
\end{eqnarray}
Equations for five parameters $\{v_0,v_1,\phi_0,e,p\}$
are given by
\begin{eqnarray}
  \label{T1}
  \frac{\partial {\cal E}}{\partial v_0} & = & 0\,\,\,\rightarrow \,\,\,
	-1+\frac{v_1}{v_0^2}\Big(\frac{e^2}{f(\phi_0)}+f(\phi_0)p^2\Big) =0,
	\\
  \label{T2}
  \frac{\partial {\cal E}}{\partial v_1} & = & 0\,\,\,\rightarrow \,\,\,
	1-\frac{1}{v_0}\Big(\frac{e^2}{f(\phi_0)}+f(\phi_0)p^2\Big)=0,
	\\
  \label{T3}
  \frac{\partial {\cal E}}{\partial \phi_0} & = & 0\,\,\,\rightarrow \,\,\,
	\Big(-\frac{e^2}{f^2(\phi_0)}+p^2\Big)f'(\phi_0) = 0,
	\\
  \label{T4}
  \frac{\partial {\cal E}}{\partial e} & = & 0\,\,\,\rightarrow \,\,\,
	Q = e \frac{v_1}{v_0}\frac{1}{ f(\phi_0)},
\\
\label{T5}
  \frac{\partial {\cal E}}{\partial p} & = & 0\,\,\,\rightarrow \,\,\,
	P = p \frac{v_1}{v_0} f(\phi_0).
\end{eqnarray}
 Summation of Eqs.\eqref{T1} and~\eqref{T2} leads to the relation
\begin{eqnarray}
v_0=v_1= \frac{e^2}{f(\phi_0)}+f(\phi_0)p^2.
\end{eqnarray}
Then,  two Eqs.\eqref{T4} and \eqref{T5} indicate
\begin{eqnarray}
Q = \frac{e}{  f(\phi_0)}, \quad P=p f(\phi_0).
\label{T6}
\end{eqnarray}
On the other hand, Eq.(\ref{T3}) implies two different conditions
\begin{equation}
f'(\phi_0)=0, \quad f(\phi_0)=\frac{e}{p}.
\end{equation}
The latter condition implies
\begin{equation}
 v_0=v_1(\equiv r_e^2)=2ep=2QP,
\end{equation}
which corresponds to scalarized extremal black hole found in Eq.(\ref{Cbh-2}) for $\phi_0=\phi_c$.
The former condition  leads to
\begin{equation}
v_0=v_1(\equiv r_e^2)=\frac{e^2}{f(\phi_0)}+f(\phi_0)p^2=Q^2f(\phi_0)+\frac{P^2}{f(\phi_0)},
\end{equation}
which is the same condition  as in Eq.(\ref{Cbh-3}) when choosing $\phi_0=\tilde{\phi}_c$.
Hence, we confirm that two scalarized extremal black hole solutions  are recovered  from the entropy function approach.

Finally, let us compute their entropy.
For an exponential coupling $f(\phi)=e^{\alpha \phi^2}$, one finds that the entropy of scalarized extremal black hole is given by the Bekenstein-Hawking entropy
\begin{equation}
\mathcal{E}_{f=e^{\alpha \phi^2}}=2\pi\Big(eQ+pP-\frac{v_1}{2}\Big)=\pi v_1=\pi (2QP)< \pi(Q^2+P^2),
\end{equation}
which shows that the entropy of scalarized extremal black hole is less than that of dyonic extrmal black hole for $Q\not=P$. 
In this case, the constant scalar hair  takes the form
\begin{equation}
\phi_0(Q,P,\alpha)=\frac{1}{\sqrt{\alpha}}\sqrt{\ln\Big[\frac{P}{Q}\Big]}, \label{sh-1}
\end{equation}
which is secondary because it is expressed in terms of all known parameters of $Q,P$, and $\alpha$.
This constant scalar hair is similar to the  horizon dilaton ($\bar{\phi}=\frac{1}{2}\ln\Big[\frac{P}{Q}\Big]$) for  dilatonic non-extremal  black hole.
Furthermore, the constant scalar in Eq.(\ref{sh-1}) is considered as a fixed scalar because its
value on the degenerate horizon is  fixed by two U(1) charges~\cite{Kol:1996hf,Krasnitz:1997gn,Lee:1997xg}.

For a polynomial form of $f(\phi)=1+\alpha \phi^2-\beta \phi^4$, one obtains  the constant scalar hair from the condition of $f'(\phi_0)=0$ as
\begin{equation}
\phi_0(\alpha,\beta)=\sqrt{\frac{\alpha}{2\beta}}, \label{sh-2}
\end{equation}
which is still secondary hair  since it is fixed by two coupling  parameters $\alpha$ and $\beta$.  For $f(\phi_0)=1+\frac{\alpha^2}{4\beta}$,
the entropy of scalarized extremal black hole is given by
\begin{equation}
\mathcal{E}_{f=1+\alpha\phi^2-\beta\phi^4}=2\pi\Big(eQ+pP-\frac{v_1}{2}\Big)=\pi v_1=\pi \Big[f(\phi_0)Q^2+\frac{P^2}{f(\phi_0)}\Big].
\end{equation}
We find two inequalities such  that $\mathcal{E}_{f=1+\alpha\phi^2-\beta\phi^4}> \mathcal{E}_{dbh}$ for $Q>P$,  while $\mathcal{E}_{f=1+\alpha\phi^2-\beta\phi^4} < \mathcal{E}_{dbh}$ for $Q<P$ with the entropy of dyonic extremal black hole $\mathcal{E}_{dbh}=\pi(Q^2+P^2)$.
Finally, considering a quadratic coupling function $f(\phi)=1+\alpha \phi^2$ used for spontaneous scalarization, one could not obtain non-zero constant scalar hair $\phi_0$. 

\section{Discussions }

It is known that one famous solution for scalarized extremal  black holes  is the  BBMB black hole found from the ECS theory~\cite{Bocharova:1970skc,Bekenstein:1974sf}.
Its  secondary scalar hair takes the form of  $\phi(r)=m/(r-m)$ with $m$ black hole mass, even though it blows up at the horizon.
One believes that it is not easy to find scalarized extremal black hole with primary scalar hair (independent scalar charge $Q_s$) because requiring  an asymptotically flat condition  change primary scalar into secondary one~\cite{Myung:2019adj,Zou:2019ays}.  Hence, it  allows us  to find scalarized extremal black hole with either secondary scalar hair or constant scalar hair.

In the present work, we obtained two scalarzied extremal black holes with constant  scalar hair in the EMS theory with two different scalar coupling functions to two U(1) fields.
It is worth noting that these solution are exactly  recovered  from the extremal scalarization  and  entropy function approaches.
We note that similar scalarized extremal black hole solutions can be found from the dyonic EMS theory with  the single scalar coupling~\cite{Astefanesei:2019pfq,Chen:2026olq}.
Actually, these will be obtained from our result  by exchanging  ``$Q\leftrightarrow P$".

Finally, we would like to mention   that it is not easy to find scalarized (extremal) black holes with primary scalar hair unless its action possesses  a shift symmetry like as an internal symmetry  in (beyond) Horndeski theory~\cite{Babichev:2013cya,Bakopoulos:2023fmv}.

 \vspace{2cm}

{\bf Funding}:  X.Y.C. is supported by the starting
grant of Jiangsu University of Science and Technology (JUST) and National
Science Foundation of China (no: W2533026).
 Y.S.M. is supported by the National Research Foundation of Korea (NRF) grant funded by the Korea government (MSIT) (RS-2022-NR069013). \\
{\bf Data Availability Statement}: The original contributions presented in this study are included in
the article. \\

\end{document}